\documentclass[preprint2,twocolumn,times,tighten]{aastex631}

\usepackage{graphicx}	
\usepackage{amsmath}	
\usepackage{multirow}
\let\tablenum\relax
\usepackage{siunitx}

\begin{document}

\title{Origin of the Multi-Phase Interstellar Medium: the Effects of Turbulence and Magnetic Field}

\email{yuehu@ias.edu; *NASA Hubble Fellow}

\author[0000-0002-8455-0805]{Yue Hu*}
\affiliation{Institute for Advanced Study, 1 Einstein Drive, Princeton, NJ 08540, USA }


\begin{abstract}
The interstellar medium (ISM) consists of a multiphase gas, including the warm neutral medium (WNM), the unstable neutral medium (UNM), and the cold neutral medium (CNM). While significant attention has been devoted to understanding the WNM and CNM, the formation of a substantial fraction of the UNM, with temperatures ranging from a few hundred to a few thousand Kelvin, remains less well understood. In this study, we present three-dimensional hydrodynamical and magnetohydrodynamical simulations of turbulent multiphase ISM to investigate the roles of turbulence and magnetic fields in regulating the multiphase ISM. Our results confirm that turbulence is crucial in redistributing energy and producing the UNM. The turbulent mixing effect smooths the phase diagram, flattens the pressure-density relationship, and increases the fraction of gas in the UNM. We find magnetic fields not only contribute to sustain the UNM but also influence the dynamics and distribution of gas across all phases. Magnetic fields introduce anisotropy to the turbulent cascade, reducing the efficiency of turbulent mixing in the direction perpendicular to the magnetic field. We find the anisotropy results in a less flat phase diagram compared to hydrodynamical cases. Furthermore, the inclusion of magnetic fields shallowens the second-order velocity structure functions across multiple ISM phases, suggesting that more small-scale fluctuations are driven. These fluctuations contribute to the formation of the UNM by altering the energy cascade and thermodynamic properties of the gas. Our findings suggest that the combined effects of turbulence and magnetic fields are important in regulating the multiphase ISM. 
\end{abstract}




\section{Introduction} \label{sec:intro}
The interstellar medium (ISM) is a highly complex environment, composed of gas, dust, and magnetic fields spanning a wide range of densities and temperatures \citep{1999MNRAS.302..417S,2000A&A...354..247L,2010ApJ...710..853C,2011piim.book.....D,2011Sci...334..955L,2012ARA&A..50...29C,2012ARA&A..50..491P,Gray2017,2019NatAs...3..776H,2022ApJ...934....7H}. Traditionally, the ISM has been modeled as consisting of multiple phases in pressure equilibrium: a warm, diffuse medium; a unstable, intermediate-density medium; and a cold, dense medium \citep{1977ApJ...218..148M,1995ApJ...443..152W,2003ApJ...587..278W,2011piim.book.....D}. While much attention has been given to both the warm and cold components \citep{1990ApJ...352L...1R,2002ApJ...564L..97K,2008A&A...486L..43H,2008A&ARv..15..189S,2021NatAs...5..365F,2021MNRAS.500.3290M}, the unstable medium (with temperatures ranging from a few hundred to a few thousand Kelvin) plays a crucial role in mediating the exchange of mass and energy between phases \citep{1965ApJ...142..531F,1989ApJ...336..808H,2001ApJ...557L.121G}. Understanding how this phase forms, survives, and evolves is critical, as it acts as a transition zone between the warm and cold ISM, influencing thermal balance, turbulence dissipation, and the efficiency of phase mixing \citep{2000ApJ...540..271V,2003ApJ...589L..77C,2023ApJ...949L...5F,2024RvMPP...8...21Y,2024arXiv240714199H}. The interplay between these phases regulates star formation, feedback, and galaxy evolution across multiple scales \citep{2017ARA&A..55..389T,2019ApJ...878..157X,2022MNRAS.514..957S}.

A key mechanism shaping the unstable phase of ISM is turbulence \citep{2000ApJ...540..271V,2001ApJ...557L.121G,2003ApJ...589L..77C,2012ApJ...747...86C,2023ApJ...949L...5F,2024RvMPP...8...21Y,2024arXiv240714199H,2025arXiv250303305C}. Observational studies have shown that turbulence is ubiquitous across the ISM and affects processes such as phase transitions, cloud fragmentation, and energy dissipation \citep{1995ApJ...443..209A,2010ApJ...710..853C,2022ApJ...934....7H}. In particular, turbulence is expected to drive mixing layers that form at the interface of cold and warm gas, populating the intermediate-temperature unstable phase \citep{1990MNRAS.244P..26B,2019MNRAS.487..737J,2020ApJ...894L..24F}. The interaction of turbulence with magnetic fields further complicates these processes: magnetic fields can induce anisotropy to turbulent flows \citep{LV99,2003ApJ...589L..77C,2021ApJ...911...37H,2024arXiv241008157H}, influencing gas dynamics and structure formation \citep{2000ApJ...537..720L,2005LNP...664..137H,2023MNRAS.524.2994H}. It is thus important to account for the combined effects of turbulence and magnetic fields in regulating the unstable phase when modeling the ISM.

Earlier numerical and theoretical work has attempted to investigate the role of turbulence in regulating the multi-phase structure of the ISM. The thermal-instability models have examined how gas fragments into two-phase medium (warm neutral medium, WNM, and cold neutral medium, CNM, see \citealt{1965ApJ...142..531F}), and 2D magnetohydrodynamic (MHD) simulations have been used to study the effect of turbulence mixing in sustaining the unstable neutral medium (UNM, see \citealt{2000ApJ...540..271V,2001ApJ...557L.121G}). While such studies of turbulence mixing effects have been extended by analyzing 3D simulations \citep{2024arXiv240714199H}, less attention was paid to the magnetic field. Especially, a full understanding of how the combined effects of turbulence and magnetic fields specifically sustain the unstable regime remains elusive. In this work, we present 3D, unmagnetized and magnetized, turbulent multi-phase ISM simulations continuously driven on large scales. Our simulations include atomic line cooling and photoelectric heating, allowing warm, intermediate-temperature, and cold gas to form. We investigate the efficiency of turbulent mixing in redistributing mass and energy and how magnetic fields mediate phase boundaries. We focus on how these processes specifically impact the survival of intermediate-temperature gas. We also investigate the MHD turbulence's properties across different phases. 

This paper is organized as follows. In Section~\ref{sec:theory}, we derive the turbulent mixing model from the ideal MHD equations and further consider the magnetic field's' effect. Section~\ref{sec:data} describes the details of the 3D simulations used in this study. In Section~\ref{sec:results}, we examine the effects of turbulence and magnetic fields by analyzing simulations with different parameters, focusing on phase diagrams and turbulence statistics. Finally, we summarize our findings in Section~\ref{sec:conclusion}.

\section{Theoretical consideration}
\label{sec:theory} 
\subsection{Turbulent mixing effect}
To study the turbulent mixing effect, we use the ideal MHD equations. The total energy equation is given by:
\begin{equation} 
\frac{\partial E}{\partial t} + \nabla \cdot \left[\pmb{v}\Bigl(E + P + \frac{B^2}{8\pi}\Bigr) - \frac{\mathbf{B}(\mathbf{B}\cdot\pmb{v})}{4\pi}\right] = \Gamma - \Lambda, 
\label{eq:total_energy} 
\end{equation} 
with the total energy density:
\begin{equation} E = \frac{1}{2}\rho v^2 + \rho\epsilon + \frac{B^2}{8\pi}.
\label{eq:Etot} 
\end{equation} 
Here, $\rho$ is the mass density, $v = |\pmb{v}|$ the gas speed, $\epsilon$ the specific internal energy, $P$ the gas pressure, and $B = |\mathbf{B}|$ the magnetic field strength; 
$\Gamma$ and $\Lambda$ denote volumetric heating and cooling, respectively. For an ideal gas (with adiabatic index $\gamma$), the internal energy is written as $\epsilon =c_vT$, where $c_v$ is the specific heat at constant volume and $T$ is the temperature.

By subtracting the kinetic and magnetic energy contributions—which satisfy:
\begin{equation} 
\frac{\partial}{\partial t}\left(\frac{B^2}{8\pi}\right) + \nabla \cdot \left[\frac{\mathbf{B}\times(\pmb{v}\times\mathbf{B})}{4\pi} - \pmb{v}\cdot\frac{(\nabla\times\mathbf{B})\times\mathbf{B}}{4\pi}\right] = 0, \label{eq:B_energy} 
\end{equation} 
and 
\begin{equation} 
\frac{\partial}{\partial t}\left(\frac{1}{2}\rho v^2\right) + \nabla \cdot \left(\frac{1}{2}\rho v^2\pmb{v}\right) = -P\nabla\cdot\pmb{v} + \frac{\pmb{v}\cdot\bigl[(\nabla\times\mathbf{B})\times\mathbf{B}\bigr]}{4\pi}, \label{eq:K_energy} 
\end{equation} 
—we obtain an equation for the evolution of the internal energy. Expressing the internal energy in terms of temperature yields:
\begin{equation} 
c_v \frac{\partial (\rho T)}{\partial t} + c_v \nabla \cdot (\rho T\pmb{v}) = -c_v (\gamma-1)(\rho T)\cdot\nabla \pmb{v} + \Gamma - \Lambda . 
\label{eq:internal_energy} 
\end{equation}

Next, we introduce a Reynolds decomposition. At each spatial point $\mathbf{x}$ we write
\[
\rho(\mathbf{x}) = \bar{\rho}(\mathbf{x}) + \rho'(\mathbf{x}), \quad
T(\mathbf{x}) = \bar{T}(\mathbf{x}) + T'(\mathbf{x}),
\]
and we assume that the velocity field is entirely turbulent (i.e., the mean flow vanishes),
\[
\pmb{v}(\mathbf{x}) = \pmb{v}'(\mathbf{x}).
\]
Here, an overbar denotes the Reynolds average over length scale $l_{\rm tur}$ and a prime the corresponding fluctuation.

Substituting these decompositions into Eq.~\eqref{eq:internal_energy} and applying the Reynolds averaging rules yields:
\begin{equation}
\begin{aligned}
  c_v\frac{\partial}{\partial t}\Bigl(\bar{\rho}\bar{T} + \overline{\rho'T'}\Bigr)
  &+ c_v\nabla\cdot\Bigl(\bar{\rho}\overline{T'v'} + \bar{T}\overline{\rho'v'} + \overline{\rho'\,T'v'}\Bigr) \\
  &= - c_v(\gamma-1)\overline{(\rho'T')\cdot\nabla \pmb{v}'} + \overline{\Gamma} - \overline{\Lambda}~.
\end{aligned}
\label{eq:reynolds}
\end{equation}
In Eq.~\eqref{eq:reynolds} the second-order correlations (such as $\overline{T'v'}$ and $\overline{\rho'v'}$) represent the turbulent fluxes of temperature and density, while the third-order term $\overline{\rho' T'v'}$ and the correlation $\overline{(\rho'T')\cdot\nabla \pmb{v}'}$ capture higher-order turbulent effects.  The third-order terms are neglected for simplicity of calculations.

Assuming a steady state (so that the time derivative vanishes) and neglecting third-order correlations, we model the turbulent fluxes by the standard gradient‐diffusion approximations:
\[
\overline{T'v'} \approx -\kappa_T\,\nabla\bar{T}, \qquad
\overline{\rho'v'} \approx -\kappa_\rho\,\nabla\bar{\rho}\,,
\]
where $\kappa_T$ and $\kappa_\rho$ are effective turbulent diffusion coefficients. Typically $\kappa_T$ and $\kappa_\rho$ are proportional to $v_{\rm tur} l_{\rm tur}$,
with $v_{\rm tur}$ and $l_{\rm tur}$ denoting a characteristic turbulent velocity and length scale, respectively. With these closures,  Eq.~\eqref{eq:reynolds} becomes
\begin{equation}
  \overline{\Gamma} - \overline{\Lambda} + \underbrace{c_v\,\kappa_T\,\nabla\cdot\Bigl(\bar{\rho}\,\nabla\bar{T}\Bigr)}_{\text{turbulent temperature mixing}} + \underbrace{c_v \kappa_\rho\nabla\cdot\Bigl(\bar{T}\nabla\bar{\rho}\Bigr)}_{\text{turbulent density mixing}} = 0\,.
  \label{eq:diffusion}
\end{equation}
Eq.~\eqref{eq:diffusion} expresses the balance between net heating/cooling and the turbulent mixing (modeled as diffusive transport) of temperature and density. It suggests that turbulent mixing weakens the tendency of thermal condensation, smoothing the phase separation. 

\subsection{Magnetic field's dual effects}
\textbf{Introduce anisotropy:} Fluctuations induced by MHD turbulence were initially considered isotropic, neglecting the influence of magnetic fields \citep{1964SvA.....7..566I, 1965PhFl....8.1385K}. However, subsequent analytical studies \citep{GS95,LV99}, numerical simulations \citep{2000ApJ...539..273C, 2001ApJ...554.1175M, 2003MNRAS.345..325C, 2010ApJ...720..742K, HXL21, 2024MNRAS.527.3945H}, and in situ solar wind measurements \citep{2016ApJ...816...15W, 2020FrASS...7...83M, 2021ApJ...915L...8D, 2023arXiv230512507Z} have established a modern understanding of MHD turbulence, characterized by scale-dependent anisotropy. This anisotropy modifies turbulent diffusion coefficients (see Eq.~\ref{eq:diffusion}).

A shift in understanding this anisotropy stems from the "critical balance" condition, first formulated by \cite{GS95} in the global mean magnetic field reference frame and later refined by \cite{LV99} in the local magnetic field reference frame, which is more appropriate for studying the local turbulent mixing effect. The "critical balance" condition posits that the turbulence cascading time is balanced by the Alfv\'en wave period \citep{LV99}: 
\begin{equation}
\label{eq.cb}
    l_\bot^{-1} v_{{\rm tur},l_\bot}\approx l_\|^{-1} v_A, 
\end{equation}
where $v_A = B/\sqrt{4\pi\rho}$ is the Alfv\'en speed, with $B$ and $\rho$ being the magnetic field and gas density, respectively, while $v_{{\rm tur},l_\bot}$ refers to the turbulent velocity at the scale $l_\bot$ measured in the local reference system, defined relative to the magnetic field intersecting an eddy. $l_\bot$ and $l_\parallel$ are the scales perpendicular and parallel to the local magnetic field, respectively. The local system of reference was explicitly defined in  \cite{2000ApJ...539..273C}. Eq.~\ref{eq.cb} is based on the idea that turbulent reconnection of magnetic field lines, occurring within an eddy turnover time, facilitates magnetic line mixing perpendicular to their orientation, thereby minimizing resistance to turbulent cascading. This process enables eddies to cascade efficiently in the direction perpendicular to the local magnetic field.
\begin{figure*}
\centering
\includegraphics[width=0.99\linewidth]{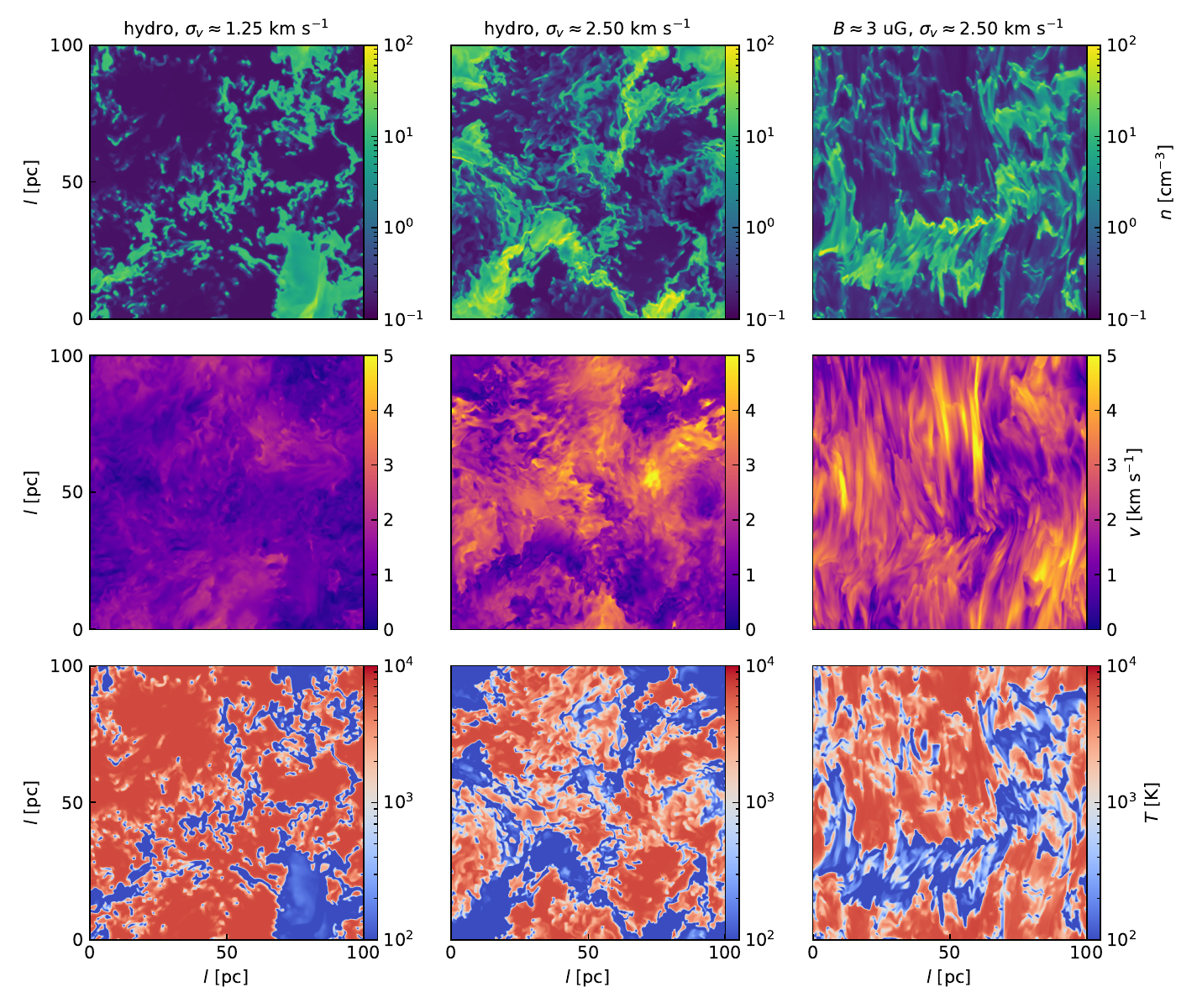}
        \caption{Slices of number density (top panel), velocity (middle panel), and temperature (bottom panel). Three different physical conditions are included: hydrodynamic cubes with $\sigma_v\approx1.25$~km s$^{-1}$ (left panel), hydrodynamic cubes with $\sigma_v\approx2.50$~km s$^{-1}$ (central panel), and MHD cubes with $\sigma_v\approx2.50$~km s$^{-1}$ (right panel). For the MHD case, the mean magnetic field is vertical.}
    \label{fig:vrho_slice}
\end{figure*}

By adapting the "critical balance" condition to the local reference frame and incorporating the Kolmogorov scaling relation $v_{{\rm tur},l_\bot} = (\frac{l_\bot}{L_{\rm inj}})^{1/3} v_{\rm inj}M_A^{1/3}$, where $M_A=v_{\rm inj}/v_A$ is the Alfv\'en Mach number and $v_{\rm inj}$ is the turbulent velocity at the injection scale $L_{\rm inj}$, one can derive the scale-dependent anisotropy scaling in the strong turbulence regime \footnote{The strong turbulence regime in sub-Alfv\'enic turbulence ($M_A<$ 1) spans from the transitional scale $l_{\rm trans} = L_{\rm inj}M^2_A$ to smaller scales. Turbulence within the range from $L_{\rm inj}$ to $l_{\rm trans}$ is termed weak turbulence, which is wave-like and does not obey the "critical balance". The weak turbulence is also anisotropic, but the scaling relation is different from Eq.~\ref{eq.lv99} \citep{2001ApJ...562..279L,xu2019study}.
For super-Alfv\'enic turbulence ($M_A>$ 1), the strong turbulence regime spans from $l_{\rm a} = L_{\rm inj}M^{-3}_A$ to smaller scales.}\citep{LV99}:
\begin{align}
\label{eq.lv99}
 l_\bot&= L_{\rm inj}(\frac{l_\parallel}{L_{\rm inj}})^{\frac{3}{2}} M_A^{2},\\
 v_{{\rm tur},l_\bot}&= v_{\rm inj}(\frac{l_\parallel}{L_{\rm inj}})^{\frac{1}{2}} M_A,
\end{align}
which reveals \textit{the anisotropic nature of turbulent eddies, with $l_\parallel \gg l_\bot$ for the same velocity fluctuation levels}.

This anisotropy directly impacts the turbulent diffusion coefficients $\kappa_T$ and $\kappa_\rho$, as the turbulent cascade is preferentially channeled in the direction perpendicular to the magnetic field. Therefore, the coefficients are given by:
\begin{align}
    \kappa \propto v_{\rm inj}l_\parallel(\frac{l_\parallel}{ L_{\rm inj}})M_A^{3},
\end{align}
indicating that turbulent diffusion is slower in a strongly magnetized medium, i.e., $M_A<1$. For weakly magnetized medium with $M_A>1$, the anisotropy is reduced and magnetic field's role is insignificant. $\kappa$ refers to either $\kappa_T$ or $\kappa_\rho$.

\textbf{Drive fluctuations and provide support:} Note that the magnetic field can have dual effects. In addition to the anisotropy, the magnetic field introduces its own fluctuations, which can alter the thermal equilibrium by inducing additional fluctuations in the total gas pressure and temperature. Magnetic pressure provides more support against compression, preventing the ISM from cleanly separating into cold, dense clouds and warm, diffuse gas \citep{2023MNRAS.521..230H}. The net effect is that, in the phase diagram, the magnetic field reduces deviations from thermal condensation while simultaneously inducing more fluctuations around the mean profile.


\begin{table}
	\centering
 \begin{tabular}{ | c | c | c | c | c | c |}
		\hline
		Run & Condition & ${\rm n}$ [cm$^{-3}$] & $\sigma_v$ [km s$^{-1}$] & $B$ [uG] &  $M_A$ \\ \hline \hline
		A0 & Hydro & 3 & 1.25 & 0 & - \\ 
		A1 & Hydro & 3 &  2.50 & 0 & - \\
            A2 & Hydro & 3 &  5.00 & 0 & - \\\hline
		A3 & MHD & 3 &  1.25 & 3.0 &  0.37 \\
            A4 & MHD & 3 & 2.50 & 3.0 & 0.75 \\
            A5 & MHD & 3 & 5.00 & 3.0 & 1.5 \\\hline
            A6 & MHD & 3 & 1.25 & 5.0 & 0.22 \\
            A7 & MHD & 3 & 2.50 & 5.0 & 0.45 \\
            A8 & MHD & 3 & 5.00 & 5.0 & 0.9 \\
        \hline
	\end{tabular}
	\caption{\label{tab:sim} Initial setups of multiphase ISM simulations. $n$, $\sigma_v$, and $B$ are the number density, velocity dispersion, and magnetic field strength, respectively.  $M_A$ represents the initial Alfv\'en Mach number.
 }
\end{table}

\section{Numerical simulations} 
\label{sec:data}

The 3D multi-phase ISM simulations analyzed in this study were generated using the AthenaK code \citep{2024arXiv240916053S}. These simulations solve the ideal MHD equations with periodic boundary conditions, given by:

\begin{equation} 
\label{eq.mhd} 
\begin{aligned} 
&\frac{\partial\rho}{\partial t} +\nabla\cdot\left(\rho\pmb{v}\right)=0,\\ 
\\
&\frac{\partial(\rho\pmb{v})}{\partial t}+\nabla\cdot\left[\rho\pmb{v}\pmb{v}^T+\left(c_s^2\rho+\frac{B^2}{8\pi}\right)\mathbf{I}-\frac{\mathbf{B}\mathbf{B}^T}{4\pi}\right] = \mathbf{f},\\ 
\\
&\frac{\partial\mathbf{B}}{\partial t} - \nabla\times(\pmb{v}\times\mathbf{B})=0,\\ 
\\
&\nabla \cdot\mathbf{B}=0,\\ 
\\
&\frac{\partial E}{\partial t} + \nabla \cdot \left[\pmb{v}\Bigl(E + P + \frac{B^2}{8\pi}\Bigr) - \frac{\mathbf{B}(\mathbf{B}\cdot\pmb{v})}{4\pi}\right] = \Gamma - \Lambda,
\end{aligned} 
\end{equation} where a stochastic forcing term, $\mathbf{f}$, is applied to drive turbulence. The cooling energy density rate, $\Lambda$, accounts for atomic line cooling and is given by \citep{2002ApJ...564L..97K}: 
\begin{equation} 
\begin{aligned} 
\Lambda = &\left(\frac{\rho}{m_{\rm H}}\right)^2\Bigl[2\times10^{-19}\exp\left(\frac{-114800}{T+1000}\right)\\ 
&+ 2.8\times10^{-28}\sqrt{T}\exp\left(\frac{-92}{T}\right)\Bigr] ~~ {\rm erg~s^{-1}cm^{-3}}, 
\end{aligned} 
\end{equation} 
where $m_{\rm H}$ is the hydrogen mass. 
The heating energy density rate, $\Gamma$, is given by: 
\begin{equation} 
\begin{aligned} 
\Gamma = & \left(\frac{\rho}{m_{\rm H}}\right)\times10^{-26} ~~ {\rm erg~s^{-1}~cm^{-3}}. 
\end{aligned}
\end{equation}
The thermal conduction term is excluded from the energy equation in order to isolate the effects of turbulence and magnetic fields. \citet{2004ApJ...602L..25K} emphasized that satisfying the Field condition via thermal conduction is essential for accurately capturing thermal instability in numerical simulations, while \citet{2000ApJ...540..271V} found that thermal instability becomes subdominant in a turbulent medium. If thermal conduction were included, we expect that the sharpness of the temperature transition between the WNM and CNM would be further smoothed, potentially increasing the fraction of the UNM.

The initial conditions include a uniform number density field of $n = 3~{\rm cm^{-3}}$ and a uniform magnetic field aligned along the $y$-axis. We choose $B \approx3$~\SI{}{\micro G} and $\approx5$~\SI{}{\micro G} based on Zeeman observation \citep{2012ARA&A..50...29C}. For hydrodynamic simulations, the magnetic field is set to zero. Three values of velocity dispersion are included: $\sigma_v\approx1.25$, 2.50, and 5.00 km s$^{-1}$. The choice of $\sigma_v\approx 5.00$ km s$^{-1}$ is based on Larson's law and observational constraints from young stars \citep{1981MNRAS.194..809L,2022ApJ...934....7H,2023AAS...24122801H}. These parameters are summarized in Table~\ref{tab:sim}. The simulation box size is 100~pc, with turbulence driven solenoidally at a peak wavenumber of 2. The computational domain is discretized on a uniform $512^3$ grid, with numerical dissipation occurring at scales of approximately 10 cells. The simulations are evolved for 100~Myr. More details on the implementation of the turbulence driving can be found in \cite{2024MNRAS.527.3945H}.

\begin{figure*}[p]
\centering
\includegraphics[width=1.0\linewidth]{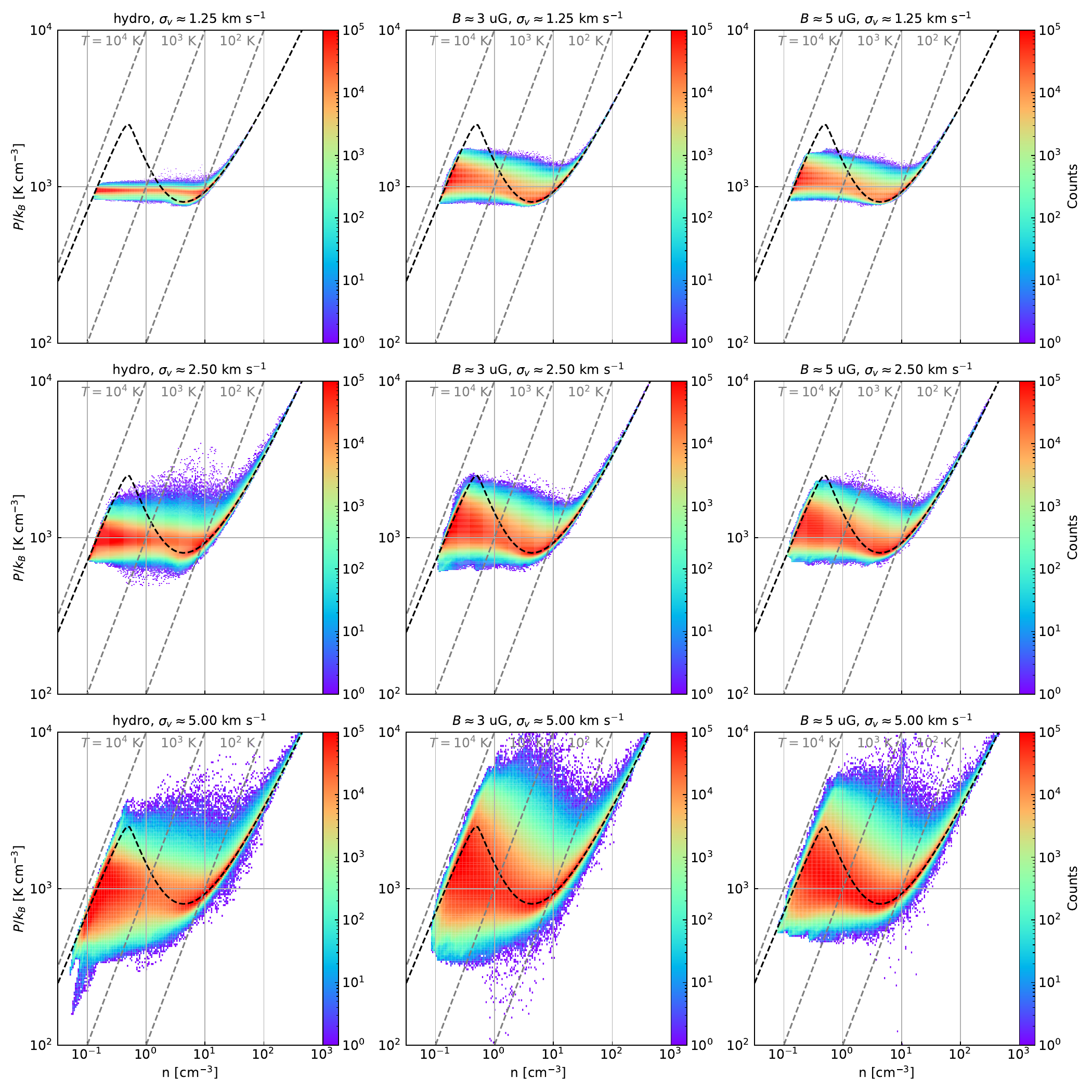}
        \caption{Phase diagrams of gas density and pressure. The black dashed line represents the thermal equilibrium obtained from $\Gamma = \Lambda$, where $\Gamma$ and $\Lambda$ are the heating and cooling functions, respectively. Three different magnetic field conditions: $B = 0$ (hydro), $\approx3$, and $\approx\SI{5}{\micro G}$, as well as three turbulence conditions: $\sigma_v\approx1.25$, 2.50, and 5.00 km s$^{-1}$ are included.}
    \label{fig:phase_diagram}
\end{figure*}
\section{Results} 
\label{sec:results}
\subsection{Turbulence induces fluctuations and magnetic field introduces anisotropy}
Fig.~\ref{fig:vrho_slice} presents the 2D slices of number density, velocity, and temperature under three distinct physical conditions: a hydrodynamic simulation with $\sigma_v\approx1.25$~km s$^{-1}$, a hydrodynamic simulation with $\sigma_v\approx2.50$~km s$^{-1}$, and an MHD simulation with $\sigma_v\approx2.50$~km s$^{-1}$ and $B\approx\SI{3}{\micro G}$. Turbulence introduces fluctuations in both number density and velocity, with the magnitude of these fluctuations increasing for higher $\sigma_v$. The spatial distributions of these fluctuations are isotropic, showing no preferential direction. For temperature fluctuations, turbulence with higher amplitude (that is, $\sigma_v\approx2.50$~ km s$^{-1}$) smooths the transition from the WNM to the CNM, resulting in less distinct boundaries and the emergence of more intermediate temperature UNM. When a magnetic field is included, the distributions of number density, velocity, and temperature become anisotropic, similar to that under isothermal conditions \citep{2010ApJ...720..742K,2020MNRAS.498.1593B,2024MNRAS.527.3945H}. Structures tend to elongate along the magnetic field direction, with this anisotropy being most pronounced in the velocity field. For density structures, both parallel and perpendicular orientations relative to the magnetic field are observed. The perpendicular structures are most likely formed by shocks. 

\subsection{Effects of turbulence and magnetic fields on the thermodynamic fluctuations}
Fig.~\ref{fig:phase_diagram} shows the phase diagrams of gas density and pressure under three different magnetic field conditions: $B = 0$ (hydro), $\approx3$, and $\approx\SI{5}{\micro G}$, as well as three turbulence conditions: $\sigma_v\approx1.25$, 2.50, and 5.00 km s$^{-1}$. For the hydrodynamic case with $\sigma_v\approx1.25$ km s$^{-1}$, the pressure remains approximately constant at $P/k_B\approx10^3~$ K cm$^{-3}$ around when the temperature exceeds 100 K, resulting in a flat phase curve that deviates significantly from the thermal condensation curve derived from $\Gamma = \Lambda$, where $\Gamma$ and $\Lambda$ represent the heating and cooling functions, respectively. This deviation arises because turbulence efficiently redistributes energy and mixes the WNM, UNM, and CNM, producing a smoother average profile in the phase diagram. However, turbulence has a dual effect on the thermodynamic state of the medium. On the one hand, it redistributes energy, flattening the phase diagram. On the other hand, turbulence induces fluctuations around this mean state, increasing the scatter of temperature, pressure, and density. As $\sigma_v$ increases to $\approx2.50$ and 5.00 km s$^{-1}$, the majority of the phase diagram remains flat due to the efficient redistribution of energy. However, the scatter in the phase diagram becomes more pronounced, reflecting the significance of turbulence.

Moreover, the magnetic field plays a crucial role in regulating thermodynamic fluctuations. As shown in Fig.~\ref{fig:phase_diagram},  the inclusion of a magnetic field causes the flat phase curve to become steeper, with higher pressure observed in the low-density WNM range. This change arises from the anisotropy introduced into the turbulence by the magnetic field. Specifically, the anisotropy reduces turbulent diffusion coefficients in the direction perpendicular to the magnetic field, thereby slowing the efficiency of turbulent mixing (see \S~\ref{sec:theory}). Similar to the hydrodynamic case, as $\sigma_v$ increases to $\approx2.50$ and 5.00 km s$^{-1}$, turbulence induces fluctuations around the mean thermodynamic state, increasing the scatter in temperature, pressure, and density. On the other hand, even if the turbulence level remains constant, the presence of the magnetic field introduces fluctuations to pressure. In addition to the driven turbulence, these magnetic fluctuations can arise from effects such as magnetic reconnection, turbulent dynamo, and other instabilities that can affect local pressure and density. These fluctuations result in larger scatter around the mean profile in the phase diagram.
\begin{figure*}
\centering
\includegraphics[width=0.99\linewidth]{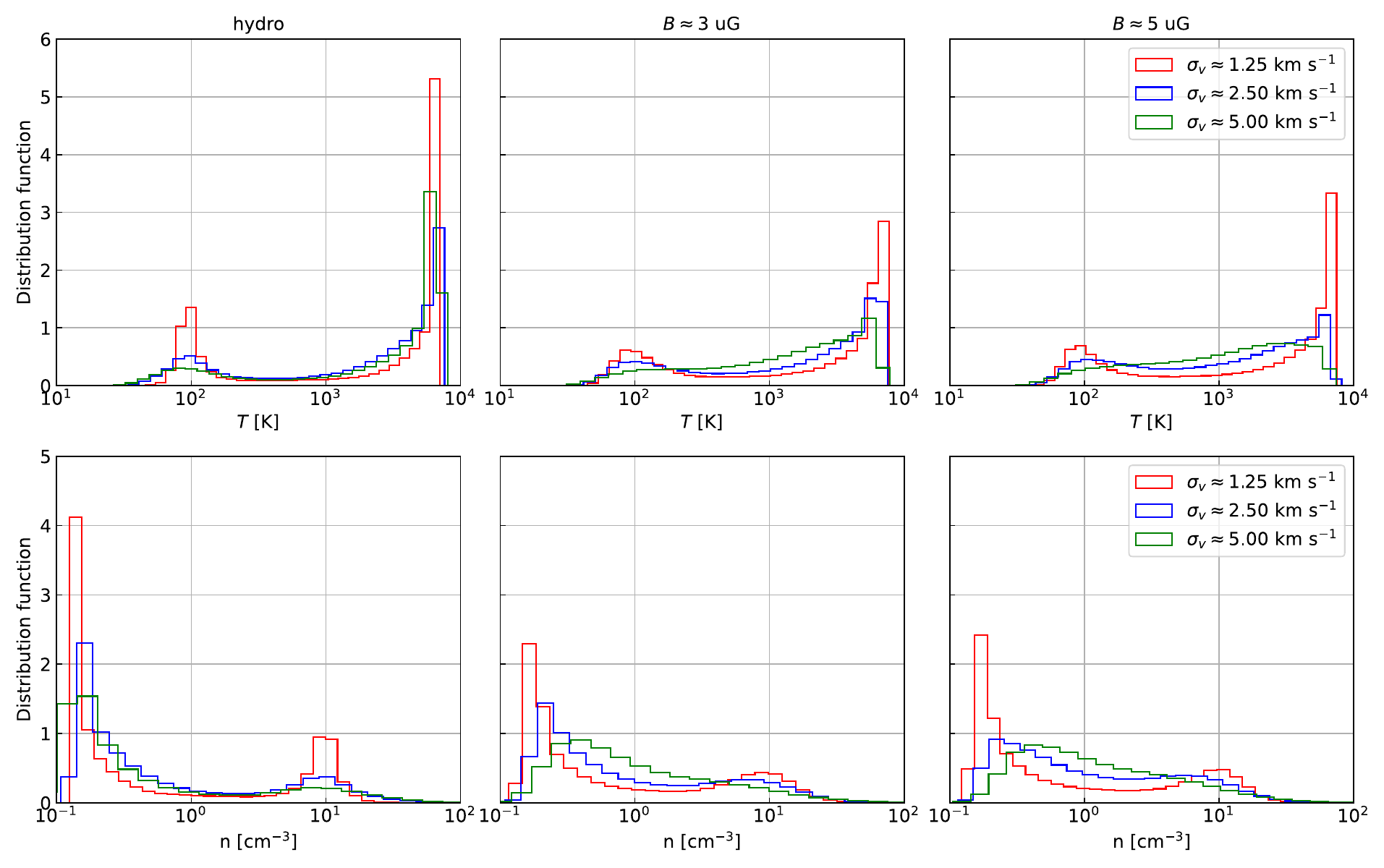}
        \caption{Histograms of gas temperature (top) and number density (bottom). 
        For comparison, three different magnetic field conditions: $B = 0$ (hydro), $\approx3$, and $\approx\SI{5}{\micro G}$, as well as three turbulence conditions: $\sigma_v\approx1.25$, 2.50, and 5.00 km s$^{-1}$ are included.}
    \label{fig:hist_nT}
\end{figure*}

\begin{figure*}
\centering
\includegraphics[width=0.99\linewidth]{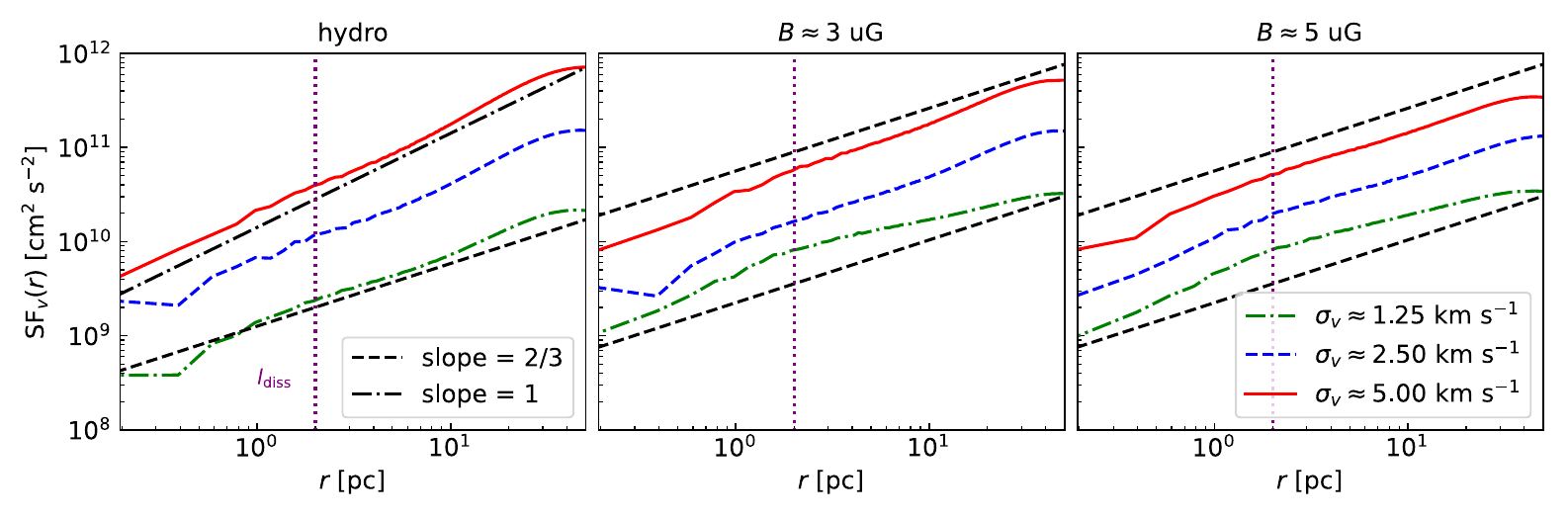}
        \caption{The undecomposed structure function of velocity. The structure function is calculated for three different magnetic field conditions: $B = 0$ (hydro), $\approx3$, and $\approx\SI{5}{\micro G}$, as well as three turbulence conditions: $\sigma_v\approx1.25$, 2.50, and 5.00 km s$^{-1}$. To guide the eye, the dashed and dash-dotted lines represent power-law slopes of 2/3 and 1, for comparison with a Kolmogorov and Burgers turbulence scaling, respectively.}
    \label{fig:sf}
\end{figure*}

\begin{figure*}
\centering
\includegraphics[width=0.99\linewidth]{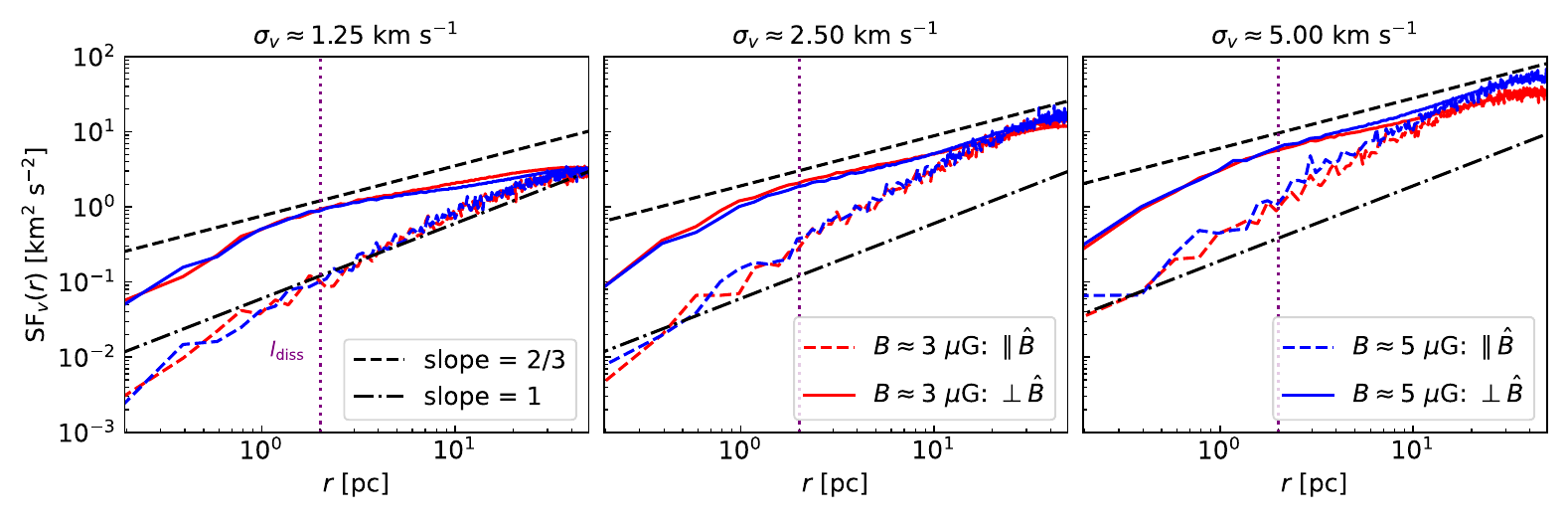}
        \caption{The parallel and perpendicular components of the decomposed velocity structure function. The decomposition is performed with respect to the local magnetic field. The structure function is calculated for three turbulence conditions: $\sigma_v\approx1.25$, 2.50, and 5.00 km s$^{-1}$. To guide the eye, the dashed and dash-dotted lines represent power-law slopes of 1/3 and 1/2, for comparison with a Kolmogorov and Burgers turbulence scaling, respectively.}
    \label{fig:sf_dec}
\end{figure*}
\subsection{Turbulence and magnetic fields sustain the UNM}
The histograms of gas temperature and number density under are shown in Fig.~\ref{fig:hist_nT}. For the hydrodynamical case with $\sigma_v\approx1.25$ km s$^{-1}$, two distinct phases, WNM and CNM, are formed at temperature $T\leq200$~K and $T\geq5000$~K. The fraction of the intermediate temperature ($T\approx200 - 5000$~K) UNM is small. The three phases have a number density around 0.1, 1, and 10 cm$^{-3}$. When $\sigma_v$ increases, the number density fraction of WNM and CNM decreases, while the UNM fraction increases due to a stronger turbulent mixing effect. 

When a magnetic field is included, the phase separation becomes smoother under the same turbulence conditions. The number density fractions of the WNM and CNM decrease further, while the fraction of gas in the UNM increases compared to the hydrodynamic cases. This behavior arises because the magnetic field introduces additional fluctuations in pressure (see Fig.~\ref{fig:phase_diagram}). For instance, the magnetic field pressure can provide support that makes the compression of gas into the CNM more difficult. While the magnetic field weakens turbulent mixing by reducing diffusion efficiency perpendicular to the field, fluctuations in magnetic field pressure sustain the UNM. As a result, instead of a sharp separation between the WNM and CNM, the ISM exhibits a more continuous distribution of temperature and density. More significant turbulence, characterized by larger $\sigma_v$, further smooths the temperature and density distributions, leading to a larger fraction of gas in the thermally unstable phase. Under realistic ISM conditions ($B = 3 - \SI{5}{\micro G}$ and $\sigma_v\approx5.00$ km s$^{-1}$ at 100~pc), the volume fraction of UNM reaches $\geq50\%$, which is consistent with previous studies \citep{2000ApJ...540..271V,2001ApJ...557L.121G,2024arXiv240714199H}.

\subsection{MHD turbulence is anisotropic}
We utilize the second-order structure function to quantify the statistical properties of turbulent flows \citep{2000ApJ...539..273C,2021ApJ...910...88X,2021ApJ...911...37H}. The structure function for velocity is defined as:
\begin{equation}
\label{eq.sf}
{\rm SF_2}(r)=\langle|\pmb{v}(\pmb{r}_1)-\pmb{v}(\pmb{r}_2)|^2\rangle_r,
\end{equation}
where $\pmb{v}(\pmb{r}_1)$ and $\pmb{v}(\pmb{r}_2)$ are the velocities at position $\pmb{r}_1$ and $\pmb{r}_2$, and $r=|\pmb{r}_1-\pmb{r}_2|$ represents the separation. To explore anisotropy in MHD turbulence, we decompose the structure-function into components parallel and perpendicular to the local magnetic fields $\pmb{B}_{\rm loc}$, following the method described in \citet{2000ApJ...539..273C}:
\begin{align}
\label{eq.sf_loc}
&\pmb{B}_{\rm loc}=\frac{1}{2}(\boldsymbol{B}(\pmb{r}_1)+\pmb{B}(\pmb{r}_2)),\\
&{\rm SF}_2^{v}(l_\bot,l_\parallel)=\langle|\pmb{v}(\pmb{r}_1)-\pmb{v}(\pmb{r}_2)|^2\rangle_r,
\end{align}
where $r=\sqrt{l_\bot^2+l_\parallel^2}$ is defined in a cylindrical coordinate system, with the $\hat{\pmb{l}}_\parallel=\hat{\pmb{B}}_{\rm loc}$ and $l_\bot=|\hat{\pmb{l}}_\parallel\times\pmb{r}|$. 


Fig.~\ref{fig:sf} shows the second-order structure function for velocity. For the hydrodynamic case with $\sigma_v\approx1.25$ km s$^{-1}$, the structure function exhibits a power-law slope of 2/3, consistent with Kolmogorov-type turbulence scaling. However, for the case of $\sigma_v\approx5.00$ km s$^{-1}$, the structure function steepens, with a power-law slope close to 1. This steepening is partially attributed to shocks in the supersonic medium, which efficiently dissipate turbulent kinetic energy. However, as shown in \S~\ref{subsec:wuc}, the steepening also happens in WNM, which is globally subsonic, suggesting that the phase transition may also change the slope.

\begin{figure*}
\centering
\includegraphics[width=0.99\linewidth]{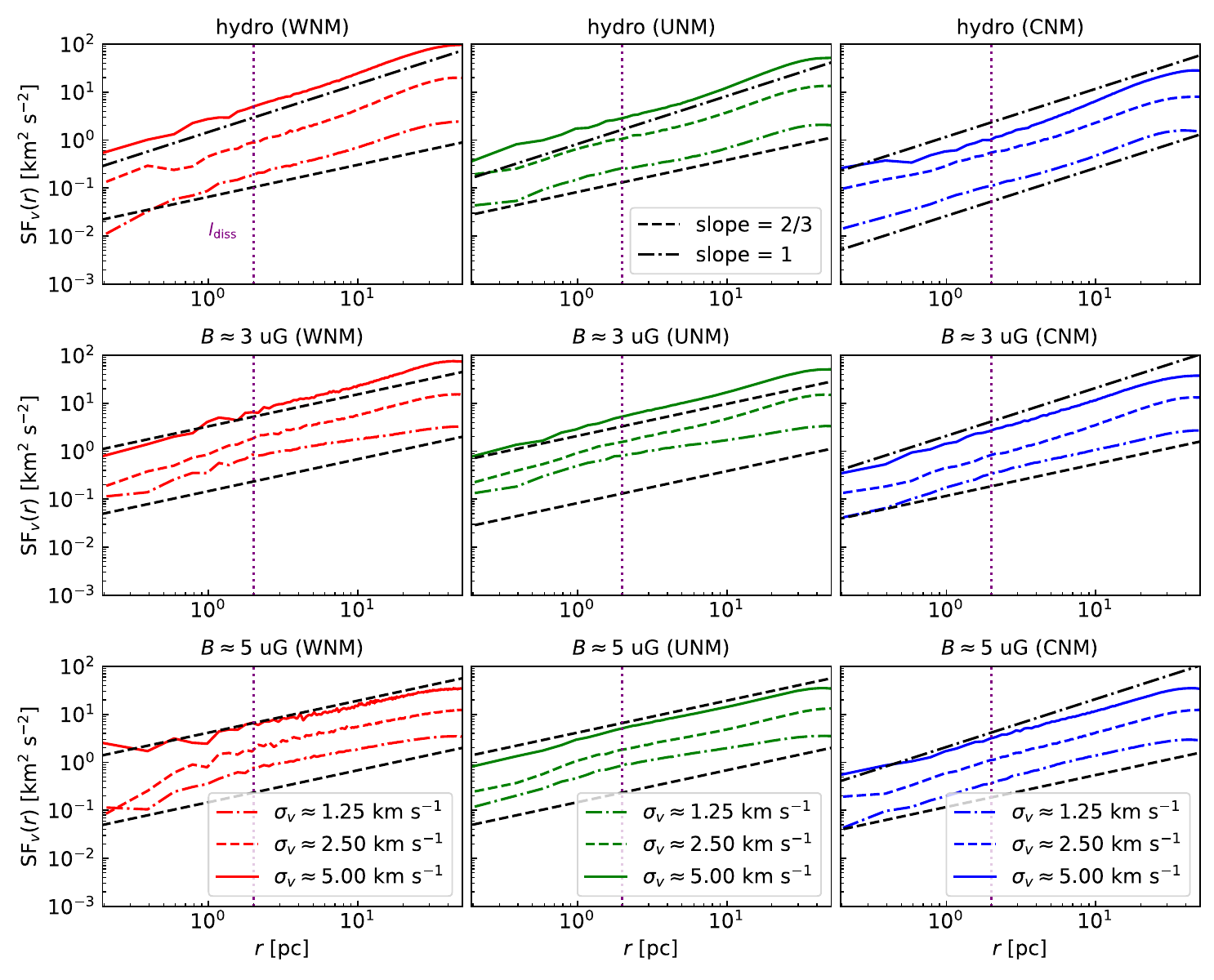}
        \caption{The structure function of velocity in WNN (i.e., $T>5000$~K), UNM (200K~$<T<5000$~K), and CNM ($T<200~$K). The structure function is calculated for three different magnetic field conditions: $B = 0$ (hydro), $\approx3$, and $\approx\SI{5}{\micro G}$, as well as three turbulence conditions: $\sigma_v\approx1.25$, 2.50, and 5.00 km s$^{-1}$. To guide the eye, the dashed and dash-dotted lines represent power-law slopes of 2/3 and 1, for comparison with a Kolmogorov and Burgers turbulence scaling, respectively.}
    \label{fig:sf_wuc}
\end{figure*}

When a magnetic field is included, the power-law slope becomes shallower than 2/3 for $\sigma_v\approx1.25$ km s$^{-1}$, indicating the presence of more small-scale fluctuations. This effect was not observed in previous isothermal MHD turbulence simulations \citep{2010ApJ...720..742K,2024MNRAS.527.3945H}, suggesting that the magnetic field introduces instabilities, which drive additional small-scale fluctuations. 
When $\sigma_v$ increases to $\approx2.50$ and $\approx5.00$ km s$^{-1}$, the structure functions remain shallower than their hydrodynamic counterparts, with power-law slopes closer to 2/3. In addition to the small-scale fluctuations driven by the magnetic field, magnetic pressure likely resists compression. As a result, the shock effects are effectively suppressed, and the turbulence retains characteristics closer to Kolmogorov scaling, even in supersonic regimes.

Fig.~\ref{fig:sf_dec} shows the decomposed second-order velocity structure function for three turbulence conditions: $\sigma_v\approx1.25$, 2.50, and 5.00 km s$^{-1}$, focusing exclusively on the MHD conditions. The power-law slope of the perpendicular component is similar to that of the undecomposed structure function (see Fig.~\ref{fig:sf}), while the parallel component exhibits a steeper slope. This indicates that the perpendicular component has a larger amplitude than the parallel component, highlighting the anisotropic nature of the MHD turbulence. Specifically, the results suggest that turbulent kinetic energy predominantly cascades in the direction perpendicular to the magnetic field in a non-isothermal medium. The ratio of the perpendicular to parallel components increases toward smaller scales, consistent with the expectation that the anisotropy strengthens at smaller scales. As discussed in \S~\ref{sec:theory}, this anisotropy suppresses the turbulence mixing effect, which is reflected in the observed shift in the phase diagram (see Fig.~\ref{fig:phase_diagram}).

\subsection{Magnetic field affects the dynamics of WNM, UNM, and CNM}
\label{subsec:wuc}
To investigate whether the magnetic field affects the turbulence's statistics in different phases, we calculate the (un-decomposed) structure function for velocity in WNN (i.e., $T>5000$~K), UNM (200K~$<T<5000$~K), and CNM ($T<200~$K). The results are shown in Fig.~\ref{fig:sf_wuc}.

For the hydrodynamical case with $\sigma_v\approx1.25$ km s$^{-1}$, the velocity structure function's power-law slope for the WNM and CNM is steeper than 2/3, approaching 1. In contrast, the UNM's slope is closer to 2/3, consistent with Kolmogorov-type turbulence. As $\sigma_v$ increases, the structure functions for the WNM, UNM, and CNM all steepen. For the UNM and CNM, the slope becomes steeper than 1. In the CNM, this steepening can be partially due to shocks, as the low temperature and low sound speed in this phase make it easier for shocks to form. However, shock dissipation typically results in a slope of 1 \citep{2010ApJ...720..742K,2024MNRAS.527.3945H}, and shocks are not expected to play a significant role in the WNM. A slope steeper than or close to 1 across all three phases suggests that additional processes, such as phase transitions or thermal condensation, may contribute to the efficient dissipation of kinetic energy.

When the magnetic field is included, the power-law slope of the structure function becomes shallower than in the hydrodynamical case across all three phases. As the magnetic field strength increases, this effect becomes more pronounced. When $\sigma_v\approx5.00$ km s$^{-1}$ and $B \approx\SI{5}{\micro G}$, the WNM and UNM exhibit slopes close to 2/3, while the CNM's slope lies between 1 and 2/3. For lower turbulence levels $\sigma_v\approx1.25$ km s$^{-1}$, the slopes are even shallower than 2/3. This behavior suggests that the magnetic field introduces instabilities and drives additional small-scale fluctuations that alter the energy cascade. The magnetic field's influence is important across all three phases.

\subsection{Velocity gradient is preferentially perpendicular to the magnetic field}

\begin{figure}
\centering
\includegraphics[width=0.99\linewidth]{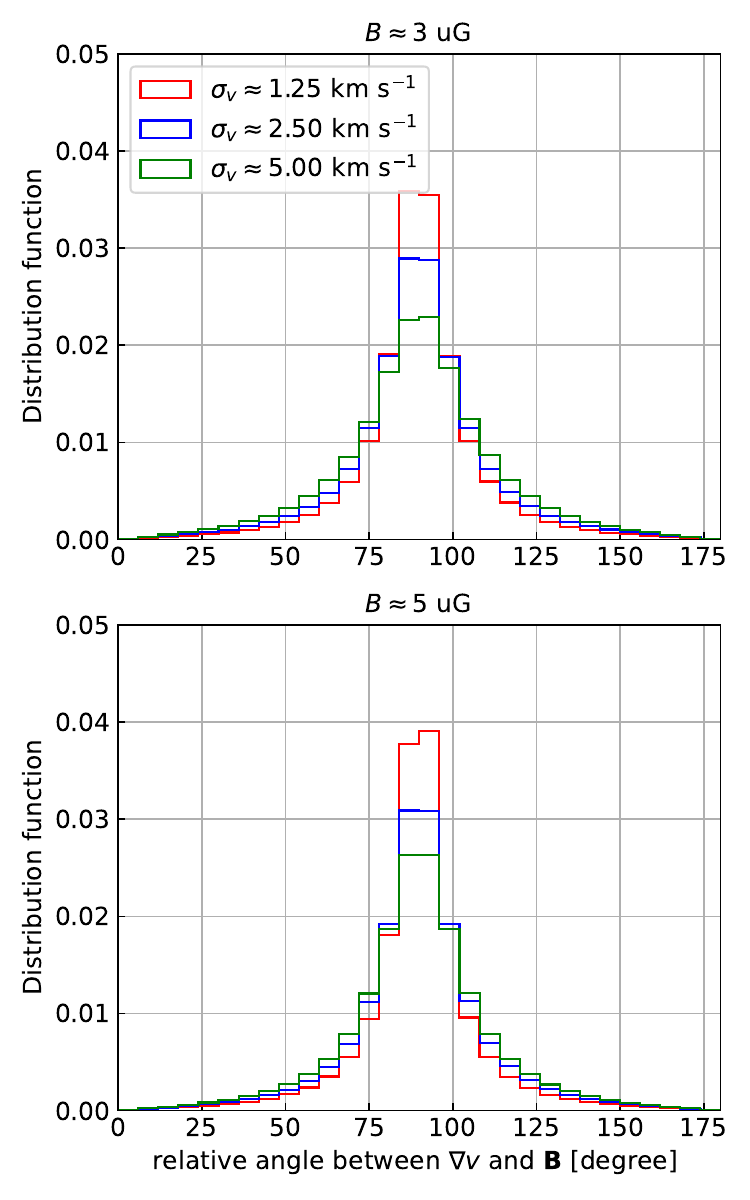}
        \caption{Histograms of the relative angle between $\nabla v$ and $\pmb{B}$ under two magnetic field conditions: $B \approx\SI{4}{\micro G}$ (top) and $\approx\SI{5}{\micro G}$ (bottom), as well as three turbulence conditions: $\sigma_v\approx1.25$, 2.50, and 5.00 km s$^{-1}$. $\nabla v$ is calculated at cell-scale $\sim0.2~$pc.}
    \label{fig:nabla v}
\end{figure}

The anisotropic nature of MHD turbulence in multi-phase ISM has important implications for studying the magnetic field. As the anisotropy suggests $l_\parallel \gg l_\bot$ (see \S~\ref{sec:theory}),  the velocity gradient can be expressed as $\nabla v_{\rm tur} \sim \frac{v_{\rm tur, l_\bot}}{l_\bot}+\frac{v_{\rm tur, l_\parallel}}{l_\parallel} \sim \frac{v_{\rm tur, l_\bot}}{l_\bot}$ meaning it is dominated by the perpendicular component. Consequently, the velocity gradient is preferentially perpendicular to the magnetic field. This anisotropy forms the foundation for using velocity gradients \citep{2017ApJ...837L..24Y,2018MNRAS.480.1333H,2018ApJ...853...96L,2020ApJ...888...96H,2022MNRAS.510.4952L}, as well as machine learning approaches \citep{2024MNRAS.52711240H} to trace magnetic field directions in the ISM.

Fig.~\ref{fig:nabla v} shows histograms of the relative angle between $\nabla v$ and $\pmb{B}$. $\nabla v$ is calculated at cell-scale $\sim0.2~$pc. Across different magnetic field strengths and turbulence conditions, the angle distributions consistently peak around 90 degrees, confirming that the velocity gradient is preferentially perpendicular to the magnetic field in the multiphase ISM. While the peak of the distribution decreases slightly as $\sigma_v$ increases, the perpendicularity remains true even under realistic ISM turbulence conditions $\sigma_v\approx 5.00$ km s$^{-1}$. 

\section{Conclusion} 
\label{sec:conclusion}
In this study, we have presented three-dimensional hydrodynamical and MHD simulations to investigate the roles of turbulence and magnetic fields in regulating the multiphase ISM. Our results demonstrate that turbulence plays a pivotal role in redistributing energy and smoothing phase transitions, thereby enhancing the fraction of gas residing in the unstable neutral medium (UNM) between the warm neutral medium (WNM) and cold neutral medium (CNM). Specifically, we find that turbulent mixing flattens the phase diagram relative to the expectations from thermal condensation.

The inclusion of magnetic fields introduces a marked anisotropy in the turbulent cascade. By reducing the effective diffusion perpendicular to the magnetic field direction and simultaneously inducing small-scale pressure fluctuations, the magnetic field modifies the thermodynamic state of the ISM. This is evidenced by a steeper phase curve and a suppression of efficient gas condensation into the CNM. This dual role of the magnetic field results in a more continuous temperature and density distribution, supporting a larger fraction of UNM. Under realistic ISM conditions ($B \approx 3–\SI{5}{\micro G}$ and $\sigma_v \approx 5$~km s$^{-1}$), the UNM comprises $\ge$50\% of the total gas, consistent with previous studies.

Furthermore, our analysis of second-order velocity structure functions reveals distinct statistical properties of turbulence in the presence of magnetic fields. In the hydrodynamic cases, steepening of the structure function at higher $\sigma_v \approx 5$~km s$^{-1}$ across WNM, UNM, and CNM, is consistent with shock dissipation and phase transitions. However, in the MHD cases, the presence of magnetic fields maintains slopes closer to 2/3, particularly in the perpendicular component, aligning with Kolmogorov scaling. This confirms the anisotropic nature of MHD turbulence in a non-isothermal medium. At small $\sigma_v \approx 1.25$~km s$^{-1}$, the inclusion of magnetic fields results in slopes shallower than 2/3, suggesting that magnetic fields introduce instabilities and drive additional small-scale fluctuations, altering the energy cascade. These statistical properties observed across the WNM, UNM, and CNM, indicate the magnetic field's effect is important across multiple phases. We confirm that the velocity gradient is preferentially perpendicular to the magnetic field in multi-phase ISM.

\begin{acknowledgments}
Y.H. thanks Alex Lazarian and James Stone for helpful discussion. Y.H. acknowledges the support for this work provided by NASA through the NASA Hubble Fellowship grant No. HST-HF2-51557.001 awarded by the Space Telescope Science Institute, which is operated by the Association of Universities for Research in Astronomy, Incorporated, under NASA contract NAS5-26555. This work used SDSC Expanse CPU and NCSA Delta CPU through allocations PHY230032, PHY230033, PHY230091, PHY230105,  PHY230178, and PHY240183, from the Advanced Cyberinfrastructure Coordination Ecosystem: Services \& Support (ACCESS) program, which is supported by National Science Foundation grants \#2138259, \#2138286, \#2138307, \#2137603, and \#2138296. 
\end{acknowledgments}

%

\vspace{5mm}

\software{Python3 \citep{10.5555/1593511}, AthenaK \citep{2024arXiv240916053S}
          }



\newpage
\bibliography{sample631}{}
\bibliographystyle{aasjournal}



\end{document}